# High-throughput Super-Resolution Imaging Chip based on Miniaturized Full-frequency Encoded-illumination

**Authors**
Xiaoyu Yang,[1,2,3] Haonan Zhang,[1] Feihong Lin,[1,2] Mingwei Tang,[1] Tawfique Hasan,[4] Clemens F. Kaminski,[3] Xu Liu[1,2]* and Qing Yang[1,2]*

**Affiliations**
1 State Key Laboratory of Extreme Photonics and Instrumentation, College of Optical Science and Engineering, Zhejiang University, Hangzhou, 310027, China
2 ZJU-Hangzhou Global Scientific and Technological Innovation Center, Zhejiang University, Hangzhou 311215, China
3 Department of Chemical Engineering and Biotechnology, University of Cambridge, Cambridge CB3 0AS, UK
4 Cambridge Graphene Centre, University of Cambridge, Cambridge CB3 0FA, UK

**Abstract**
A miniaturized full-frequency encoded illumination (mini-FEI) chip is presented for high-throughput super-resolution imaging using the spatial frequency shift (SFS) effect. A tunable full SFS scheme is achieved through propagating and evanescent wave. The multi-illumination modes are precisely and flexibly modulated by an encoded LED array. The light travels to the sample via a set of prisms, producing the super-resolution images with high signal-to-noise ratio (SNR). Mini-FEI super-resolution imaging reaches a resolution of 333 nm (~$\lambda$/4NA), close to the theoretical limit, while maintaining a large field of view (FOV) of ~1 mm$^2$. The method is validated on label-free samples including USAF Target, Star Target, and onion root tip cells, all of which could be successfully reconstructed. Through the introduction of integrated LED arrays for evanescent wave excitation, expensive laser systems can be avoided and the system significantly miniaturized. The mini-FEI super-resolution imaging chip is simple and cost effective to fabricate and can be used in conjunction with any inverted brightfield microscope frame and thus has great potential for widespread use in scientific and industrial research environments.

**Introduction**

Super-resolution microscopy overcomes the resolution limit imposed by optical diffraction which hampers traditional microscopy techniques [1]. It has become an important tool for in key discoveries in fields as diverse as biology, materials science and medical research. Various fluorescence-based super-resolution microscopic techniques, such as stochastic optical reconstruction microscopy (STORM) [2-9], stimulated emission depletion (STED) microscopy [10-14] and structured illumination microscopy (SIM) [15-19] have been reported. However, these technologies rely on fluorescent labels, which are prone to photobleaching and causing cellular phototoxicity. Furthermore, these technologies are not suitable for many non-biological samples such as nanomaterials and integrated devices.

For these reasons, label-free super-resolution microscopy techniques have been developed, including the spatial frequency compression (SFC) method utilizing hyperlenses [20–22] or microspheres [23,24], as well as the spatial frequency shift (SFS) method using waveguides [25], and others. Among these, the SFS method relies on large wave-vector illumination, shifting high spatial frequencies of the object into the passband of the traditional imaging system, thus relaying subwavelength spatial information to the far-field image sensor [26]. The SFS method is thus capable of performing undistorted and high-speed super-resolution imaging.

Fourier ptychographic microscopy (FPM) is typically used for super-resolution imaging via the SFS

effect [27-32]. Zheng et al. first proposed FPM [27], using an LED array to illuminate the sample at different angles to improve the spatial resolution. However, the maximum SFS achievable in FPM is theoretically limited by the wave-vector illuminating the sample. This is geometrically restricted in FPM with the illumination wave-vector propagating in free space. As a result the space bandwidth product (SBP) is limited. Improvements in spatial resolution are possible if evanescent fields are used providing larger wave-vectors to the sample. This led to the design of on-chip SFS structures, including nanowire ring [26], OLED luminous films [33], and wafers with tunable gratings [34]. However, However, some of these technologies may suffer from the problem of frequency deficiency, because the effective spatial frequency spectrum achievable contains empty regions as the SFS extends far beyond the cut-off spatial frequency (SF) of the microscope objective lens. Furthermore, these on-chip SFS super-resolution imaging technologies feature a small FOV, which is also a limiting factor in the achievable SBP. Another feature of current on-chip SFS technologies is the requirement for a laser source for illumination. This leads to complex external optical setups negating the advantages of on chip integration, and hinders miniaturization of instruments and cost reduction [26, 33, 34].

In this work, we present a miniaturized full-frequency encoded-illumination (mini-FEI) super-resolution imaging chip. An SFS method is proposed which fills the entire k-vector space without leaving empty regions in the super-resolution OTF. This is achieved through provision of omnidirectional SF components which are created by both freely propagating and confined evanescent waves. The chip contains an encoded LED array, through which the multiple illumination modes can be precisely modulated. The proposed mini-FEI super-resolution imaging system features an SBP increased by an order of magnitude and an SNR enhanced fivefold over existing on-chip SFS imaging technologies. Furthermore, the mini-FEI super-resolution imaging makes use of integrated LEDs to excite the evanescent field for SFS super-resolution imaging. The technology is plug-and-play and obviates the need for complex and expensive external optical set-ups. The unit is portable and compatible with conventional microscopes. The unit is compact and simple to fabricate, requiring no nanofabrication facility or knowhow, thus significantly reducing cost and offering opportunity for mass production. We show examples of the technology through application for the imaging of both fabricated nano-structures and biological samples, and discuss application prospects in biomedical research and materials science.

**Results and Discussion**

In conventional optical imaging systems, the objective lens can only receive sample information contained in SF lower than cut-off frequency ($f_0 = NA/\lambda$), where *NA* is the numerical aperture of the objective lens, and the $\lambda$ is the wavelength of illumination light. The objective lens can thus be regarded as a low-pass filter, which limits the spatial resolution of the imaging system. In the SFS method, high SF object information can be shifted into the passband of the imaging system by using illumination with a wave-vector $k_s$. In our imaging system, $k_s$ is defined as

$$k_s = 2\pi f_s = 2\pi \frac{n_{eff}}{\lambda}$$

where $f_s$ is the corresponding SFS depth provided by different illumination modes, $n_{eff} = n \cdot sin\theta$ is the effective refractive-index for individual illumination modes in the waveguide, $n$ is the refractive-index of the waveguide materials, the $\theta$ is the oblique incident angle (the angle between the incident ray and the surface normal). Therefore, $f_s$ can be denoted as

$$f_s = \frac{n \cdot sin\theta}{\lambda}$$

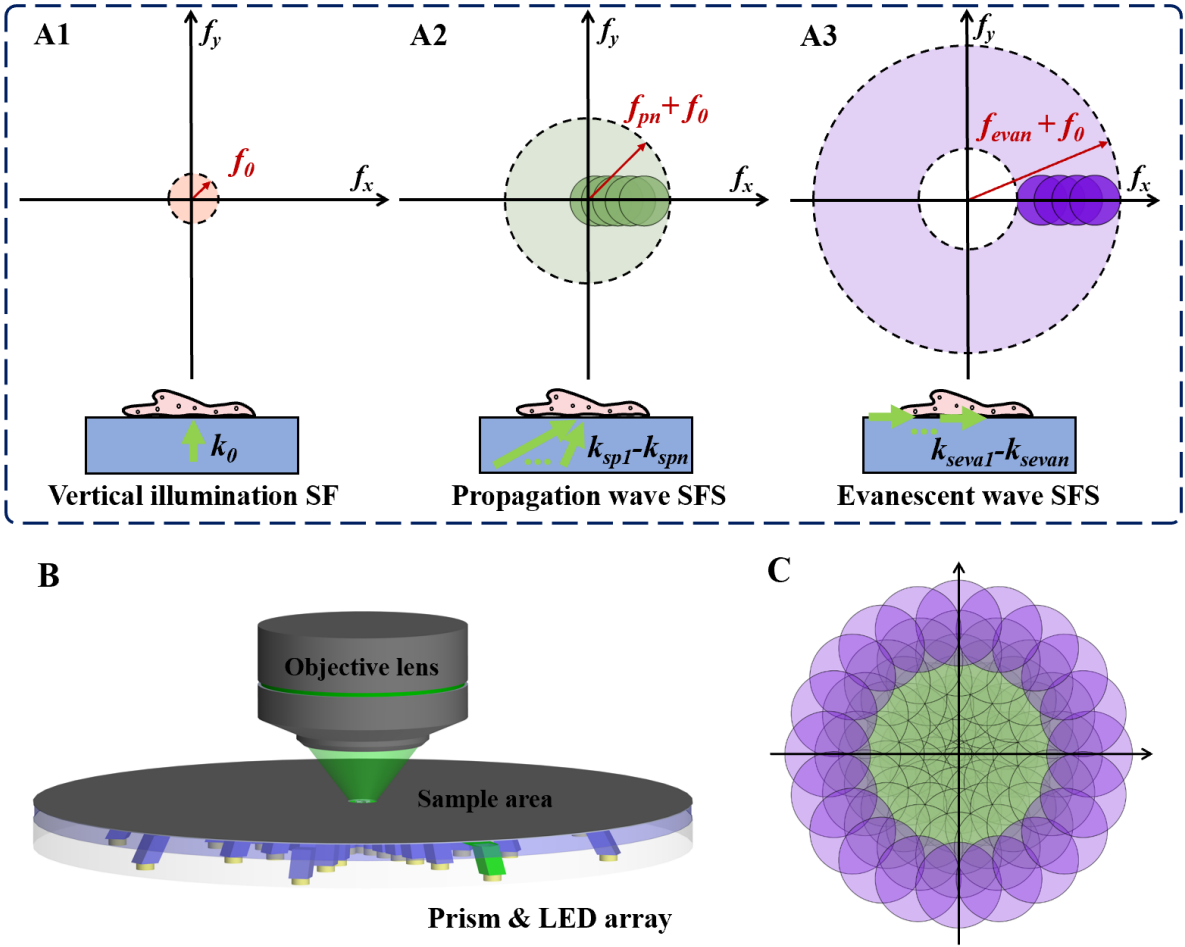

**Fig. 1. Principle of full-frequency SFS.** (A1–A3) Spatial frequencies contributed by the wave-vector under three different illumination modes: (A1) vertical illumination, (A2) oblique illumination via propagating waves, and (A3) evanescent wave illumination. (B) 3D schematic of the mini-FEI super-resolution microscopy imaging chip. (C) Corresponding spatial frequency spectrum captured using the mini-FEI imaging chip.

When illuminating objects with free space illumination in air ($n = 1$), the $f_s$ is limited to the range of $0 \leq f_s \leq 1/\lambda$, which limits the spatial resolution of the SFS imaging system. When illuminating objects in the waveguide with $n > 1$, the SFS $f_s$ can surpass $1/\lambda$, leading to a further resolution improvement, and $f_s$ can be continuously shifted to cover all frequency regions without leaving gaps.

Three illumination modes can be distinguished:

a. When $\theta = 0$, the object is vertically illuminated with wave-vector $k_0$. The SFS is $f_s = 0$, and the corresponding range of SF information obtained by SFS super-resolution imaging process is defined as $0 < f_{sr} < f_0$, as depicted in the Fig. 1(A1).
b. When $0 < \theta < \theta_c$, the object is obliquely illuminated with wave-vector $k_{spi}$ ($1 \leq i \leq n$), which is provided by a wave propagating in the waveguide. Here $0 < f_{si} < 1/\lambda$, and the corresponding range of super-resolved SF information obtained is defined as $0 < f_{sr} < f_{pn} + f_0$, as depicted in the Fig. 1(A2).
c. When $\theta \geq \theta_c$, the object is obliquely illuminated with wave-vector $k_{sevai}$ ($1 \leq i \leq n$), which is provided by evanescent waves in the waveguide. Now $f_{si} \geq 1/\lambda$, and the corresponding range of SF information spans $f_{eva1} - f_0 < f_{sr} < f_{evan} + f_0$, as depicted in the Fig. 1(A3). $\theta_c$ is the critical angle of total internal reflection.

We define $\Delta f_s = f_{s\_i+1} - f_{s\_i} = (n \cdot (\sin\theta_{i+1} - \sin\theta_i))/\lambda$, $i$ is the index of the shift frequency. Once $\Delta f_s < 2f_0$, our method has the ability to continuously adjust the SFS so that the full SF spectrum can be covered from which a super-resolution image can be reconstructed in real space without distortion (for algorithmic details, see note S1). For good performance, the overlap between shifted spectra should exceed 20% (see Note S2 and Fig.S2), so that $\Delta f_s < 1.38 f_0$ (i.e. $n \cdot (\sin\theta_{i+1} - \sin\theta_i) \leq 1.38 NA$). In contrast to FPM using free space illumination [27], our method uses waveguides to provide both propagating and evanescent wave illumination, which can overcome the limit of SFS in FPM of $1/\lambda$ to further improve the spatial resolution. Compared to SFS super-resolution imaging technology where only evanescent waves are used for illumination [26], our method leaves no gaps in the SF spectrum which would not lead to missing information and reconstruction artefacts. Moreover, we discuss the effect of using incoherent LEDs for the reconstruction algorithm in Note S3, where it is shown that despite their wide spectrum, the LEDs can be modeled as single wavelength sources in the mini-FEI algorithm.

A schematic representation of the mini-FEI super-resolution imaging chip, the corresponding spatial frequency spectrum, and experimental setup are shown in Fig. 1(B), Fig. 1(C) and Fig.S5, respectively. The illumination light from different LED sources travels through prisms onto the sample area. As the distance of the prism from the center increases, the illumination angle incident on the sample increases, which in turn increases the depth of the SFS provided by propagating and evanescent waves. The oblique incident angle is set as $\theta_i$. The wave-vector produced by the oblique incident light is $k_i = k_0 * n_{neff\_i}$. $k_0$ is the wave-vector of incident light in vacuum, which can be expressed as $k_0 = 2\pi/\lambda$. With a maximum incident angle of 53°, this super-resolution imaging system can realize a theoretical maximum resolution of 327 nm (~$\lambda$/4NA). The FOV of the imaging system is the smaller of the illuminated sample area (5 mm diameter) and the FOV of objective lens (lens dependent; 1.1 mm for the stated lens). Here an overall FOV of 1.1mm diameter is hence achieved. These parameters yield a SBP of 34.3 megapixels (MP), representing at least an order-of-magnitude improvement over previous work (see Table S2). In FPM, illumination at large incident angles with incoherent LEDs leads to very low signal because only a small fraction of the emitted light reaches the sample [27]. Chip based SFS has been proposed to partially compensate for this problem through use of gratings, which use first order diffraction to divert the incident light to the sample area at the correct angle. Modest increases in SNR have been reported in this way, however, the first-order diffraction efficiency of gratings used was relatively low (about 20%), and gains were marginal [34]. Here we use prisms with reflecting films to reflect the vertical illumination from different LEDs to the sample area. The reflection efficiency is more than 90%, which increases the SNR in collected raw images. This work demonstrates for the first time the use of encoded LEDs to excite the evanescent field for on-chip SFS super-resolution imaging, replacing conventional laser-based excitation systems which are complex and bulky to deploy. This innovation significantly miniaturizes the illumination module of super-resolution microscopes. Furthermore, the mini-FEI imaging chip eliminates the need for complex micro/nanofabrication processes, offering strong potential for low-cost industrial production.

During imaging process, when a simulated wide field image of the USAF Target vertically illuminated is obtained, which cannot distinguish any group of line pairs (Fig.2(B)). This illumination corresponds to standard brightfield imaging with the associated extent of the OTF indicated in the inset of Fig.2(B). To modulate multiple illumination modes, the LEDs are controlled by the encoded matrix shown in the bottom of Fig.2(A). The LEDs are sequentially switched from central to the outer circle, with continuously increasing SFSs including $k_0$, $k_{sp1}$, $k_{sp2}$, $k_{sp3}$, $k_{sp4}$, $k_{sp5}$, and $k_{seva1}$. To cover the entire SF spectrum for reconstruction, the larger SFSs (large $\theta_i$) require finer steps between different orientations $\varphi_j$. Therefore, for each illumination circle, the LEDs are switched from $\varphi_j = 0°$ to $360°$ with the steps of $90°$, $45°$, $30°$, 22.5°, 22.5°, 18°, respectively. This means that the number of LEDs in each circle is 4, 8, 12, 16, 16, and 20, respectively (See Fig. S6). Turning on these LEDs one by one based on the encoded set, an image stack containing high SF information of the object is obtained. As shown in top of Fig. 2(A), each SFS raw image can distinguish or enhance corresponding groups of lines. Using the image stack and the wide field image under vertical illumination, an expanded spatial frequency spectrum is iteratively reconstructed in the Fourier domain before being transformed to the spatial domain by an inverse Fourier transform, yielding a super-resolution image. A flow chart to illustrate the reconstruction algorithm is shown in Fig. S7. The theoretical resolution of the imaging system is given by $\Delta_{xy} = \lambda/(NA + k_{smax}/k_0)$, where $k_{smax}$ is the maximal wave-vector. The period of -1-6 group of lines of the simulated USAF Target is 327 nm, which is at the theoretical resolution for the mini-FEI system. Comparing the spatial frequency spectrum of the wide field image under vertical illumination with the reconstruction image, it can be found that the spatial frequency spectrum of the reconstruction image has expanded by 4 times, which is the key to resolving the nanostructures of objects beyond the diffraction limit. It indicates that our mini-FEI SFS super-resolution microscopy achieves 4-fold higher resolution than traditional wide field microscopy under identical conditions.

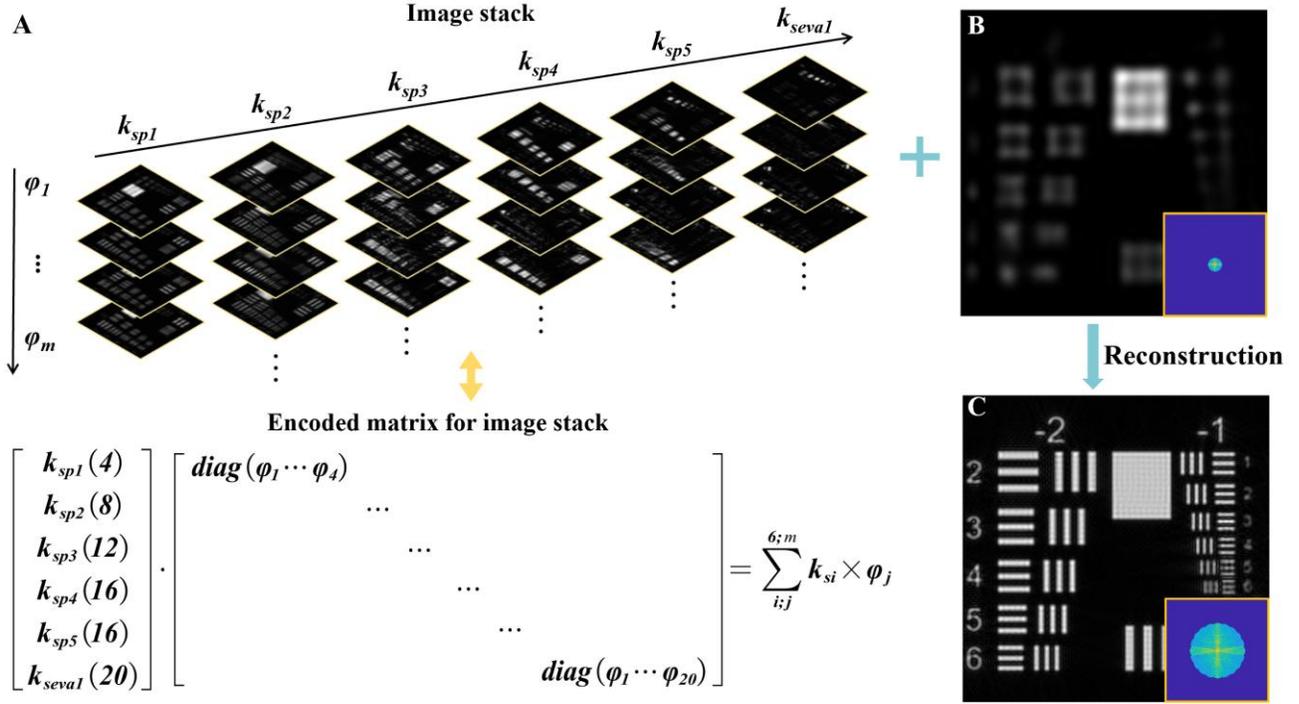

**Fig.2. Imaging process of mini-FEI super-resolution microscopy.** (A) Image stack acquired under multiple illumination angles (top), with the corresponding encoded matrix used for reconstruction (bottom). (B) Wide-field image captured under vertical illumination. The inset displays its spatial frequency spectrum. (C) Reconstructed super-resolution image. The inset shows its enhanced spatial frequency spectrum.

To validate our imaging method experimentally, a high-resolution USAF Target was used as the object. The object has a series of five-line patterns with different resolutions of 600 to 3300 lp/mm. The wide field image under vertical illumination with a ×20/0.4 NA objective lens cannot resolve frequencies beyond 700 lp/mm (corresponding to a spatial period of 1429 nm), as shown in Fig. 3(A). The finest structure resolvable with the mini-FEI super-resolution imaging system corresponds to a spatial period of 333 nm (the position of 3000 lp/mm), close to the theoretical limit. The wide field image under vertical illumination of 3000 lp/mm position is shown in Fig. 3(B), which cannot be resolved at all. By using the mini-FEI, the five-line patterns with period of 333 nm can be successfully reconstructed, as shown in Fig. 3(C). A scanning electron microscopy (SEM) image of the object (Fig. 3(D)) was obtained as the ground truth (GT). The normalized intensity profiles of the reconstructed (light red) and SEM images (light blue) are plotted in Fig. 3(E), which show good agreement with one another.

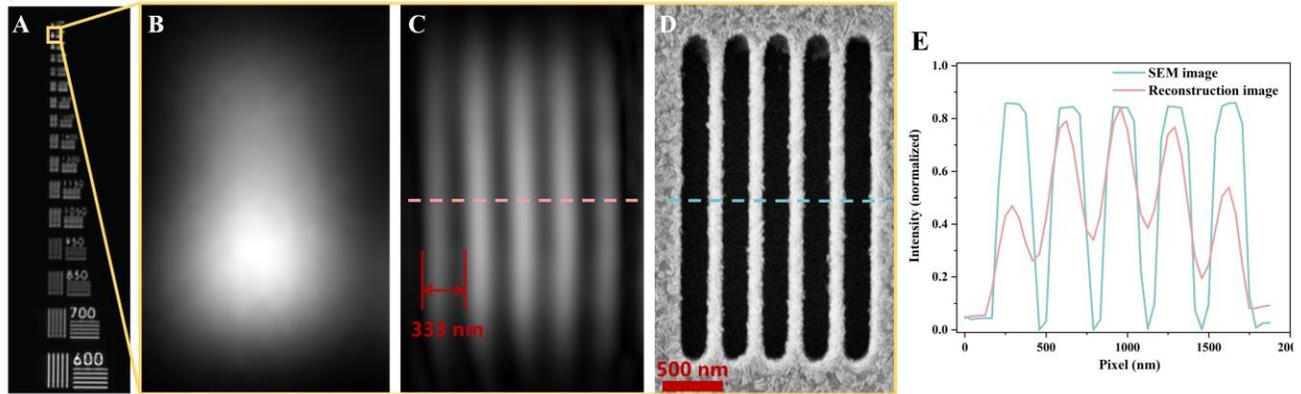

**Fig. 3. Mini-FEI Super-Resolution Imaging of Five-Line Pattern Object.** (A) Wide-field image of the five-line pattern under vertical illumination. (B) Wide-field image of the five-line pattern with 333 nm line spacing under vertical illumination. (C) The reconstruction image of five-line patterns with 333 nm resolution. (D) SEM image serving as the ground truth. (E) Normalized intensity profiles extracted along the dashed light red and light blue lines in (C) and (D), respectively, comparing the reconstructed and SEM images.

In order to demonstrate that the method provides isotropic resolution in all directions, targets with more diverse SF information were used rather than vertical and horizontal line patterns. A Star Target was used and raw images acquired by different illuminations are shown in Fig. 4(A). Each SFS (i.e. $k_{sp1}$, $k_{sp2}$, $k_{sp3}$, $k_{sp4}$, $k_{sp5}$, and $k_{seva1}$) probes different parts of the SF in the sample and different $\varphi_j$ probe frequencies across corresponding orientations. As shown in Fig. 4(B), the central of the Star Target cannot be resolved under vertical illumination (wide field) illumination. The reconstructed image, Fig. 4(C), offers much better resolution, but reconstruction artefacts are visible compared to the SEM image (Fig. 4(D)), which serves as the ground truth. The artefacts stem from the fact that the SNR in the raw images degrade for the largest SF shifts, which causes errors during reconstruction. Nevertheless, the fine structure of the Star Target is clearly resolved.

We next applied the method to image biological samples. First, images were acquired of the root tip cells of onions. To improve the image contrast, cell sections were stained using the standard method with methylrosanilinium chloride solution. The wide field image captured by the ×20/0.4NA objective lens under vertical illumination is shown in Fig. 5(A), with a corresponding FOV of ~ 1mm$^2$. Two regions of interest (ROIs) within the wide field are shown in Fig. 5(B1) and (C1). Chromosomes present in the ROIs appear blurred. The same ROIs were also captured by a ×100/1.49 NA oil immersion objective on a color CMOS camera. These super-resolution images were used as ground truth for comparison (Fig. 5(B3) and (C3)). The corresponding SFS reconstructed super-resolution images obtained with the ×20/0.4NA objective lens are shown in Fig. 5(B2) and (C2). For better presentation and easier comparison, the SFS images were mapped onto the same color scale as the GT images. Compared with wide field images under vertical illumination, the reconstructed images reveal finer details of the chromosomes, matching those of the GT images. There are five stages of mitosis in these cells, including interphase, prophase, metaphase, anaphase, and telophase. The intensity, phase and 3D images of these mitosis stages were successfully reconstructed, as shown in Fig. 5(D1-H1), Fig. 5(D2-H2) and 5(D3-H3), respectively. The 3D reconstruction images indicate the thickness distribution of sample. Compared with 2D intensity reconstruction images, the 3D reconstruction images reveal that the chromosomes are twisted across multiple planes. For instance, it can be clearly observed that two chromosomes are spatially twisted in metaphase, as marked by yellow arrow in Fig. 5(F3). To verify the correction of our method, the oil objective lens of ×100/1.49 NA was used to detect the chromosomes at different focal planes within the depth of field, see Fig.S8. It denotes that 3D reconstruction images can uncover precise and comprehensive information of the sample.

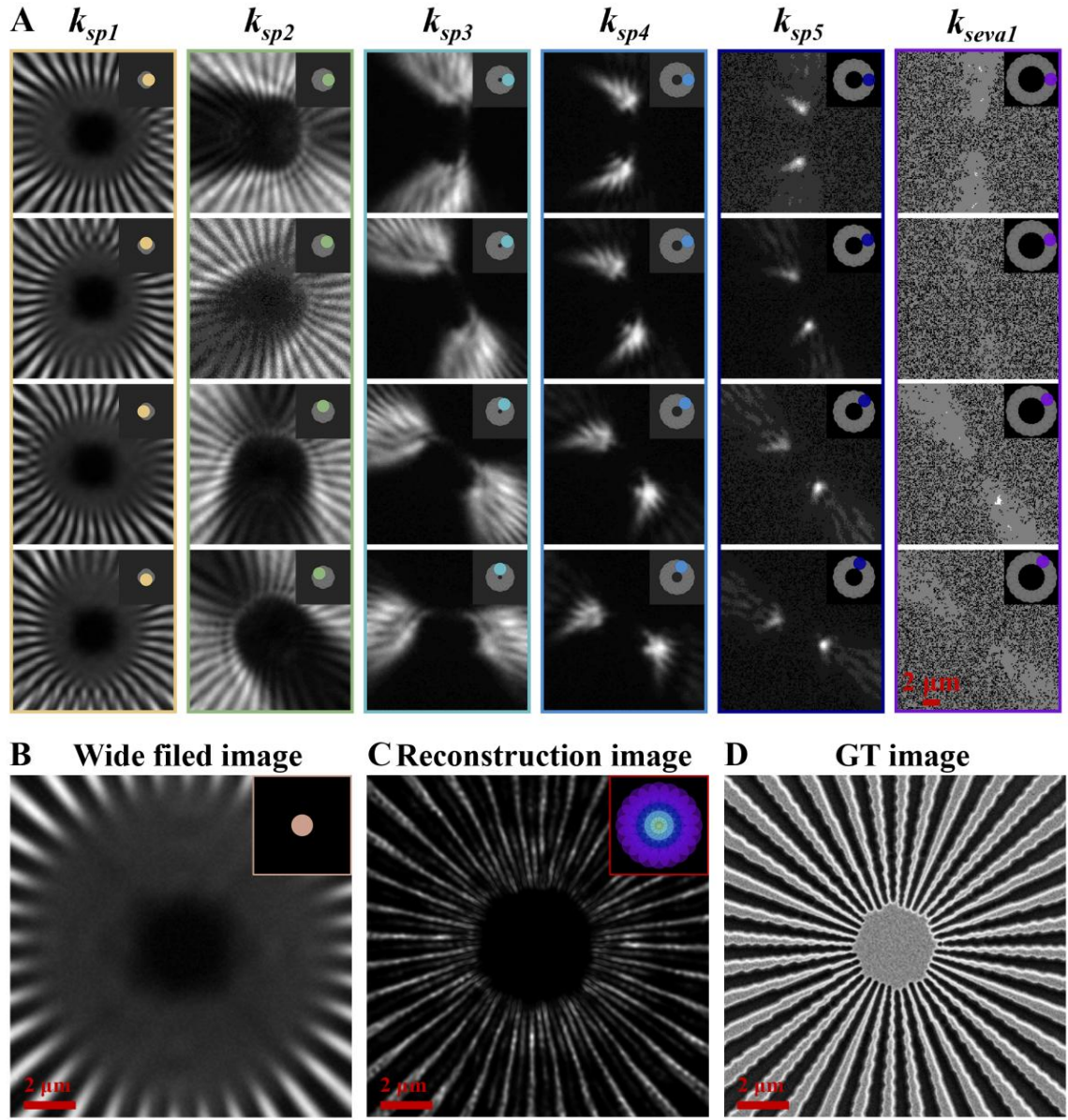

**Fig. 4. Mini-FEI Super-Resolution Imaging of a Star Target Object.** (A) Selected SFS images captured under various illumination modes, with corresponding spatial frequency spectra shown as insets. (B) Wide-field image acquired under vertical illumination, with the associated spatial frequency spectrum illustrated in the inset. (C) Reconstructed super-resolution image, accompanied by its corresponding spatial frequency spectrum (insets). (D) Scanning electron microscope (SEM) image used as the ground truth (GT). Scale bar: 2 μm.

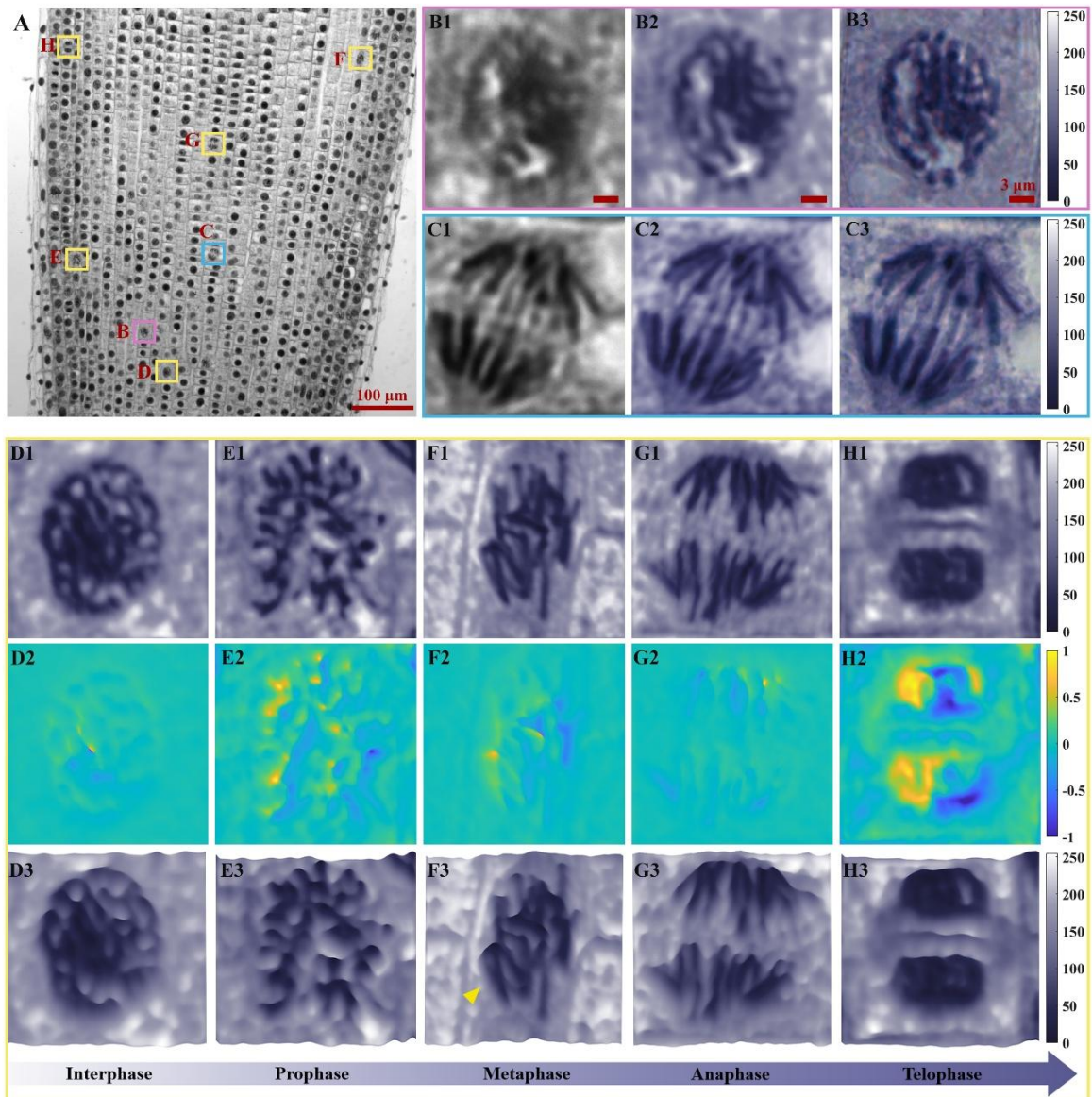

**Fig. 5. Mini-FEI Super-Resolution Imaging of Onion Root Tip Cells** (A) Wide field image acquired by a 20x/0.4NA objective under vertical illumination, showing the entire FOV of approximately 1 mm². Scale bar: 100 μm. (B1, C1) Wide field images under vertical illumination of selected ROIs. (B2, C2) Intensity reconstruction images of the corresponding ROIs. (B3, C3) GT)images of the ROIs, captured using a color CMOS camera with a 100×/1.49 NA oil-immersion objective. (D1–H1) Intensity reconstruction images of onion root tip cells at various stages of mitosis. (D2–H2) Corresponding phase reconstruction images of the same mitotic stages. (D3–H3) Three-dimensional reconstruction images showing mitotic progression. Scale bars for images (B) through (H): 3 μm.

**Conclusion**

In this work, a high-throughput integrated mini-FEI super resolution imaging system is presented. The system illuminates the sample with a combination of free space propagating and evanescent waves to permit a tunable SFS to be achieved covering the full SF spectrum of the super-resolved object. The multiple illumination modes are precisely modulated on the chip by an encoded matrix. LED light sources are Arduino controlled and delivered to waveguides via prisms integrated on the device. The prisms allow more than 90% of the LED emission to be delivered to the sample area improving the SNR of the obtained images. The imaging system realizes a large SBP of 34.3 MP with a maximal resolution of 333 nm (~$\lambda$/4NA). The mini-FEI super-resolution chip utilizes LEDs instead of lasers to excite the evanescent field, marking the first use of LEDs for this purpose. This innovation enables super-resolution imaging while significantly reducing the size of the system. The system is easy to use and compatible with standard microscope frames. No micro/nano manufacturing processes are involved in its production, significantly reducing cost and demonstrating potential for scale up and industry use. To validate performance, we tested the mini-FEI super-resolution imaging chip with a USAF Target for maximum spatial resolution, a Star Target for resolving complex structures, and onion root tip cells to demonstrate its potential for advanced applications such as digital pathology and neurogenetics.

The current mini-FEI super-resolution imaging chip operates only under an objective lens with a NA of 0.4 or greater, which typically features small FOVs. By slightly compromising the compactness, one could design a super-resolution imaging system for lower-NA objective lenses to achieve larger FOVs whilst maintaining the imaging resolution. Theoretically, higher SBPs on the order of gigapixels should be achievable. Furthermore, our current design features a refractive-index for the waveguide of 1.52, which limits the attainable resolution improvement. If materials with higher refractive index are used, further resolution gains can be achieved. Our current super-resolution imaging system only integrates the illumination module. It is possible to integrate the detection module also to miniaturize the entire SFS system, with potential for use in personal medicine and field exploration.

**Materials and Methods**
**Experiment setup**
The mini-FEI super-resolution microscopic imaging system is depicted in Fig. S5. The light source are ordinary LEDs with a center wavelength of 520 nm. A ×20/0.4NA air objective lens was used (Sanfeng). A bandpass filter with a center wavelength of 520 nm and FWHM (full width at half maximum) of 20 nm was used to improve the SNR. The imaging sensor is a commercial complementary metal oxide semiconductor (CMOS) camera with pixel size of 2.2 μm.
**Fabrication of mini-FEI super-resolution imaging chip**
The material of the waveguide is K9 glass with a refractive index of 1.52 at 520 nm. The waveguide has two parallel surfaces. The upper side serves as the imaging surface, the center of which is used as the sample holding area. To improve the SNR, all areas on the upper surface except the sample area are covered with opaque black paint. The lower surface is connected with a series of prisms arranged in concentric circles with the sample area at their center. The prism distribution is shown in Fig. S6. A full list of parameters for the mini-FEI imaging chip are shown in Table S1. LEDs are aligned vertically beneath the prism bases, supported by a 3D-printed holder. The LEDs are connected to a PCB board, which is controlled by an Arduino microcontroller. The control software for the LEDs is preloaded onto Arduino chip which makes the entire FEI system plug-and-play.
**Simulation**
All simulations were conducted in MATLAB 2023b with home-built algorithms. Additional details on the physical model simulation and mini-FEI super-resolution imaging reconstruction are provided in Note S5.

**References**
1. E. Abbe, Beiträge zur Theorie des Mikroskops und der mikroskopischen Wahrnehmung. *Arch. Mikrosk. Anat.* **9**, 413–418 (1873).
2. M. J. Rust, M. Bates, X. Zhuang, Sub-diffraction-limit imaging by stochastic optical reconstruction microscopy (STORM). *Nat. Methods* **3**, 793–795 (2006).
3. S. A. Jones, S.-H. Shim, J. He, X. Zhuang, Fast, three-dimensional super-resolution imaging of live cells. *Nat. Methods* **8**, 499–U96 (2011).
4. P. S. Probes, Multicolor super-resolution imaging with photo-switchable fluorescent probes. *Science* **317**, 1749–1753 (2007).
5. M. J. Rust, M. Bates, X. Zhuang, Sub-diffraction-limit imaging by stochastic optical reconstruction microscopy (STORM). *Nat. Methods* **20**, 231–239 (2023).
6. B. Huang, W. Wang, M. Bates, X. Zhuang, Three-dimensional super-resolution imaging by stochastic optical reconstruction microscopy (STORM) using astigmatic optics. *Science* **379**, eabq6912 (2023).
7. L. Gu, Y. Li, S. Zhang et al., Molecular resolution imaging by repetitive optical selective exposure. *Nat. Methods* **16**, 1114–1118 (2019).
8. F. Huang, G. Sirinakis, J. Bewersdorf, Quantitative analysis of molecular organization in cells using SMLM and machine learning. *Nat. Biotechnol.* **41**, 1456–1468 (2023).
9. E. Betzig, G. H. Patterson, R. Sougrat et al., High-speed 3D single-molecule tracking in living cells using interferometric SMLM. *Nat. Methods* **20**, 1124–1132 (2023).
10. S. W. Hell, J. Wichmann, Breaking the diffraction resolution limit by stimulated emission: Stimulated-emission-depletion fluorescence microscopy. *Opt. Lett.* **19**, 780–782 (1994).
11. G. Vicidomini, P. Bianchini, A. Diaspro, STED super-resolved microscopy. *Nat. Methods* **15**, 173–182 (2018).
12. F. Göttfert, C. A. Wurm, V. Mueller et al., Coaligned dual-channel STED nanoscopy and molecular diffusion analysis at 20 nm resolution. *Biophys. J.* **105**, L01–L03 (2013).
13. S. W. Hell, J. Wichmann, Breaking the diffraction resolution limit by stimulated emission depletion (STED) microscopy. *Nat. Methods* **20**, 321–334 (2023).
14. G. Vicidomini, P. Bianchini, A. Diaspro, STED super-resolution microscopy: Advances and biomedical applications. *Sci. Adv.* **9**, eadf3702 (2023).
15. M. G. L. Gustafsson, Surpassing the lateral resolution limit by a factor of two using structured illumination microscopy. *J. Microsc.* **198**, 82–87 (2000).
16. E. N. Ward, L. Hecker, C. N. Christensen et al., Machine learning assisted interferometric structured illumination microscopy for dynamic biological imaging. *Nat. Commun.* **13**, 7836 (2022).
17. E. Mudry, K. Belkebir, J. Girard et al., Structured illumination microscopy using unknown speckle patterns. *Nat. Photonics* **6**, 312–315 (2012).
18. D. Li, L. Shao, B. C. Chen et al., Extended-resolution structured illumination imaging of endocytic and cytoskeletal dynamics. *Science* **349**, aab3500 (2015).
19. L. Jin, B. Liu, F. Zhao et al., Deep learning enables structured illumination microscopy with low light levels and enhanced speed. *Nat. Commun.* **11**, 1934 (2020).
20. Z. Liu, H. Lee, Y. Xiong et al., Far-field optical hyperlens magnifying sub-diffraction-limited objects. *Science* **315**, 1686 (2007).
21. J. Rho, Z. Ye, Y. Xiong et al., Spherical hyperlens for two-dimensional sub-diffractional imaging at visible frequencies. *Nat. Commun.* **1**, 143 (2010).
22. Z. Jacob, L. V. Alekseyev, E. Narimanov, Optical hyperlens: Far-field imaging beyond the diffraction limit. *Opt. Express* **16**, 2124–2132 (2008).
23. H. Yang, N. Moullan, E. Auksorius, L. Li, Super-resolution biological microscopy using virtual imaging by a microsphere nanoscope. *Nanoscale* **7**, 15019–15026 (2015).
24. X. Hao, C. Kuang, X. Liu et al., Microsphere based microscope with optical super-resolution capability. *Appl. Phys. Lett.* **99**, 203901 (2011).
25. D. Ye, M. Tang, X. Liu et al., Low loss and omnidirectional Si3N4 waveguide for label-free spatial frequency shift super-resolution imaging. *J. Phys. D Appl. Phys.* **54**, 315101 (2021).
26. X. Liu, C. Kuang, X. Hao et al., Fluorescent nanowire ring illumination for wide-field far-field subdiffraction imaging. *Phys. Rev. Lett.* **118**, 076101 (2017).

27. G. Zheng, R. Horstmeyer, C. Yang, Wide-field, high-resolution Fourier ptychographic microscopy. *Nat. Photonics* **7**, 739–745 (2013).
28. A. Pan, C. Zuo, B. Yao, High-resolution and large field-of-view Fourier ptychographic microscopy and its applications in biomedicine. *Rep. Prog. Phys.* **83**, 096101 (2020).
29. J. Xu, T. Feng, A. Wang et al., Fourier ptychographic microscopy with adaptive resolution strategy. *Opt. Lett.* **49**, 3548–3551 (2024).
30. Y. Fan, J. Sun, Y. Shu et al., Efficient synthetic aperture for phaseless Fourier ptychographic microscopy with hybrid coherent and incoherent illumination. *Laser Photonics Rev.* **17**, 2200201 (2023).
31. Y. Shu, J. Sun, J. Lyu et al., Adaptive optical quantitative phase imaging based on annular illumination Fourier ptychographic microscopy. *PhotoniX* **3**, 24 (2022).
32. J. Chen, A. Wang, A. Pan et al., Rapid full-color Fourier ptychographic microscopy via spatially filtered color transfer. *Photonics Res.* **10**, 2410–2421 (2022).
33. C. Pang, J. Li, M. Tang et al., On-chip super-resolution imaging with fluorescent polymer films. *Adv. Funct. Mater.* **29**, 1900126 (2019).
34. M. Tang, Y. Han, D. Ye et al., High-refractive-index chip with periodically fine-tuning gratings for tunable virtual-wavevector spatial frequency shift universal super-resolution imaging. *Adv. Sci.* **9**, 2103835 (2022).

**Acknowledgments**
We thank Zhao Jiayang for suggesting that we use root tip cells of onions as the samples, which promotes the success of our experiment.

**Funding:**
National Natural Science Foundation of China (62020106002)
National Natural Science Foundation of China (92250304)
National Natural Science Foundation of China (T2293751)
National Key Research and Development Program of China (2021YFC2401403)

**Competing interests:** Authors declare that they have no competing interests.

**Data and materials availability:** All data are available in the main text or the supplementary materials.

# Supplementary Materials for

## High-throughput Super-Resolution Imaging Chip based on Miniaturized Full-frequency Encoded-illumination

Xiaoyu Yang,[1,2,3] Haonan Zhang,[1] Feihong Lin,[1,2] Mingwei Tang,[1] Tawfique Hasan,[4] Clemens F. Kaminski,[3] Xu Liu[1,2]* and Qing Yang[1,2]*

**This PDF file includes:**

> Supplementary Notes S1 to S7
> Figs. S1 to S8
> Tables S1 to S2

**Supplementary Note 1: The reconstruction images based on different illumination modes.**

To validate the advantages of full SFS illumination, we simulated four illumination modes, including vertical illumination, oblique illumination provided by propagating wave, surface illumination provided by evanescent wave and full SFS illumination. As shown in Fig. S1(A1), the fine structures of object beyond the diffraction limit cannot be resolved under the vertical illumination. As shown in Fig. S1(A2), the slots with the highest resolution cannot be resolved under oblique illumination provided by propagating wave with $k_{sp1}$-$k_{sp5}$. As shown in Fig. S1(A3), the object is almost failed to be reconstructed, and the finest structures with the high resolution are slightly reconstructed under surface illumination provided by evanescent wave with $k_{seva1}$. This is caused by the lack of the SF between the cut-off SF and the SF provided by evanescent wave, as shown in Fig. S1(B3). As shown in Fig. S1(A4), all the structures of the object are reconstructed successfully without any distortion, driven by full SFS illumination mode. Comparing with the spatial frequency spectra of illumination modes provided by propagating wave (Fig. S1(B2)) and evanescent wave (Fig. S1(B3)), the spatial frequency spectrum of the illumination modes provided by full SFS (Fig. S1(B4)) is expanded with no gap between different SF. Overall, full SFS illumination mode not only can improve the resolution compared with illumination mode provided by propagating wave, but also avoid the image distortion caused by illumination mode provided by evanescent wave.

**Supplementary Note 2: The effect of different overlap rates between SFS depths on the quality of reconstruction image.**

Different overlap rates between SFS depths influence the quality of reconstruction images. Here, the root-mean-square (RMS) error between reconstruction images and ground truth image is used to value the quality of reconstruction images. A smaller RMS error indicates better quality of the reconstructed images. As plotted in Fig.S2(B), with the increase of overlap rate from 5% to 20%, the RMS error drops steeply from 0.68 to 0.31. When the overlap rate is more than 20%, the RMS error fluctuates between 0.24~0.34. As shown in Fig. S2(A), the overlap rate is below 20%, the reconstruction images are severely distorted, while the overlap rate exceeds 20%, the reconstruction images have almost no artifacts. Overall, the overlap rates between SFS depths should be set to more than 20%.

**Supplementary Note 3: The effects of incoherent LEDs on reconstruction images.**

In the mini-FEI super-resolution imaging, the LED is regarded as the ideal coherent light source with the wavelength of 530 nm as shown in Fig. S3(B1). Its corresponding reconstruction image is shown in Fig. S3(B2), which is successfully recovered. However, the real LED used in our experiment is incoherent light source with relatively wide spectrum, whose wavelength is from 510 nm to 550 nm, centered on 530 nm, shown in Fig. S3(A1). The reconstruction image illuminated by the LEDs with wide spectrum but recovered using single central wavelength of 530 nm is shown in Fig. S3(A2), which is similar to the reconstruction image in Fig. S3(B2). It indicates that the LEDs with wide spectrum can be approximated as the LEDs with single central wavelength.

Furthermore, the central wavelength may fluctuate between 510 nm and 550 nm, influencing the quality of reconstruction images, which is displayed in Fig. S4. The LEDs used in experiment have different central wavelength between 510 nm and 550 nm, while the wavelength of 530 nm was used in the recovery process. The RMS error is firstly decreased as the wavelength is close to the 530 nm, then the RMS error is risen as the wavelength is away from the 530 nm. Notably, the RMS error is almost all below the 40% and the worst RMS error is 10% higher than RMS error in wavelength of 530 nm, which are in the acceptable range. As illustrated in the insert of Fig. S4, the finest structures of the object illuminated by LEDs with wavelengths of 510 nm and 550 nm can both be nearly successfully reconstructed, though with slight distortion, compared with those illuminated by an LED with a wavelength of 530 nm.

**Supplementary Note 4: The details of mini-FEI super-resolution microscopic system.**

The mini-FEI super-resolution microscopic system is shown in Fig. S5(A). Among this, the physical diagram and cross-section view of the mini-FEI super-resolution imaging chip are shown in Fig. S5(B) and Fig. S5(C), respectively. The K9 glass serves as the waveguide, connected with prisms and LEDs with a 3D printed holder. Then the LEDs are connected with a PCB controlled by Arduino. The thickness of waveguide is $T$. The distance between the center and the prisms in each circle is $L_i$, with different incident angles to the sample area ($\theta_i = $ arctan $(L_i/T)$), which produces different $k_{si}$. The detailed distribution of prisms is shown in Fig. S6. The detailed incident angle $\theta_i$, effective refractive-index $n_{eff}$, overlap rate between SFS depths, the number of prisms in each circle and the distance between center to prisms in each circle $L_i$ are shown in Table 1.

## Supplementary Note 5: Physical model and recovery process of mini-FEI super-resolution imaging.

In the mini-FEI super-resolution imaging system, the transverse wave-vector of illumination field generates the SFS, which is expressed as

$$E(\vec{r}_{x,(i,\varphi)}) = \exp(ik_0 n \sin\theta_{i,\varphi})$$

Suppose the complex amplitude of objects illuminated by $E(\vec{r}_{x,(i,\varphi)})$ can be denoted as $O(\vec{r}_x)$. So that the final scattering field travelling into the objective lens can be calculated as

$$U(\vec{r}_{x,(i,\varphi)}) = O(\vec{r}_x) E(\vec{r}_{x,(i,\varphi)}) = O(\vec{r}_x) \exp(ik_0 n \sin\theta_{i,\varphi})$$

Then, the process that the complex light field $U(\vec{r}_{x,(i,\varphi)})$ is collected by objective lens is regarded as convolution process through a linear time-invariant system with transfer function $H(\vec{r}_x)$. Then the light field containing the complex information of object propagates to the plane of image sensor, which can be expressed as

$$G(\vec{r}_{x,(i,\varphi)}) = U(\vec{r}_{x,(i,\varphi)}) \otimes H(\vec{r}_x)$$

Finally, the image sensor records the intensity of the above light field, denoted as

$$I(r_{i,\varphi}) = |G(\vec{r}_{x,(i,\varphi)})|^2 = |U(\vec{r}_{x,(i,\varphi)}) \otimes H(\vec{r}_x)|^2$$

In order to analyze the SFS under different illumination modes, the Fourier transform is conducted to use as the initial input of the iteration algorithm.

$$F(f_{i,\varphi}) = \mathcal{F}\{I(r_{i,\varphi})\} = \mathcal{F}\left\{|O(\vec{r}_x)\exp(ik_0 n \sin\theta_{i,\varphi}) \otimes H(\vec{r}_x)|^2\right\}$$

The recovery process of mini-FEI super-resolution imaging is illustrated in Fig. S7. The pseudo code of recovery algorithm of the mini-FEI super-resolution imaging is shown in the following:

---
**Mini-FEI Recovery Algorithm**
---
Input: $I(r_{i,\varphi}), F(f_{i,\varphi})$
Output: $I(r), p(r)$

1   for $i = 1$ to $n$ do
2      for $\varphi = 1$ to length (encoded matrix diag) do
3          $\Delta f = k_0 n \sin\theta_{i,\varphi}$
4          $F(f) \leftarrow F(f_{i,\varphi} + \Delta f)$
5          $I(r) \leftarrow abs(\mathcal{F}^{-1}\{F(f)\})$
6          $p(r) \leftarrow angle(\mathcal{F}^{-1}\{F(f)\})$

7 Output: $I(r), p(r)$

**Supplementary Note 6: 3D reconstruction images containing information of different focus planes.**

    The 3D reconstruction image of metaphase of mitosis stage in the root tip cells of onions is shown in Fig. S8(A), observing that it contains information at different focus planes. For instance, the area marked by yellow line has different chromosomes in 3D reconstruction image, which can be verified from GT images obtained by objective lens with 1.49 NA at different focus planes, shown in Fig. S8(B1) and Fig. S8(B2). It can be seen that the information in 3D reconstruction image is nearly corresponding to the information in GT images at different focus planes.

**Supplementary Note 7: Comparison of various super-resolution imaging technologies.**

Based on SFS principle, the SBP of the mini-FEI super-resolution imaging can reach to $3.4\times10^7$, which is at least an order of magnitude higher than the others. Moreover, comparing with other SFS super-resolution microscopy, the mini-FEI super-resolution microscopic chip is easy and cheap to fabricate without any complex and costly microfabrication process. The details are shown in Table S2.

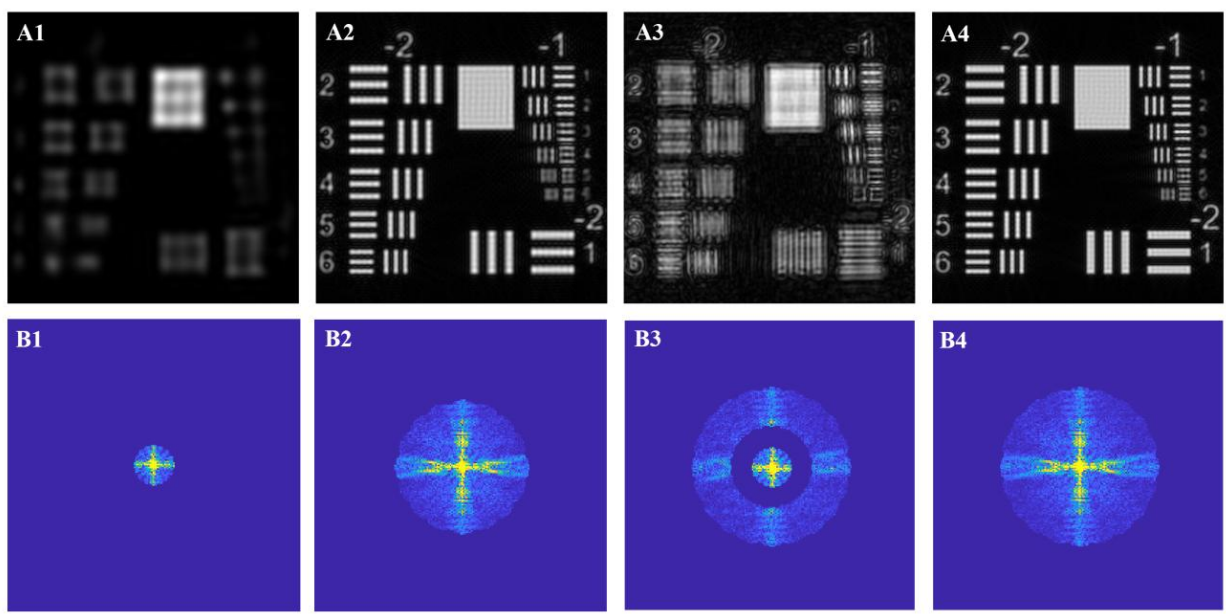

**Fig. S1. The analysis of reconstruction images based on different illumination modes.** (A1) The wide field image and (B1) the spatial frequency spectrum under vertical illumination. (A2) The reconstruction image and (B2) the spatial frequency spectrum illuminated by propagating wave with $k_{sp1}$-$k_{sp5}$. (A3) The reconstruction image and (B3) the spatial frequency spectrum illuminated by evanescent wave with $k_{seva1}$. (A4) The reconstruction image and (B4) the spatial frequency spectrum illuminated by propagating wave and evanescent wave.

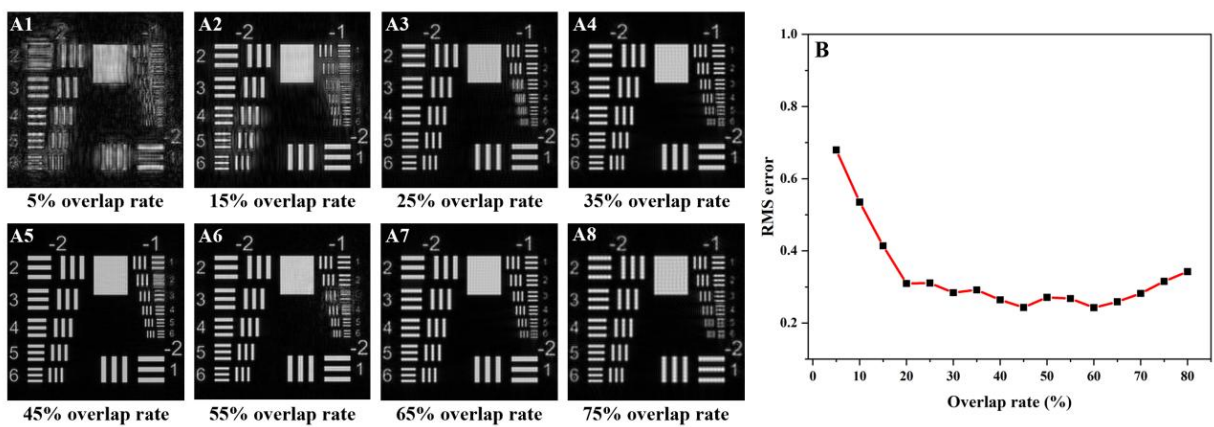

**Fig. S2. The effect of different overlap rates between SFS depths on the quality of reconstruction images.** (A) The reconstruction images with different overlap rates between SFS depths. (B) The RMS error of mini-FEI super-resolution imaging chip with different overlap rates between SFS depths.

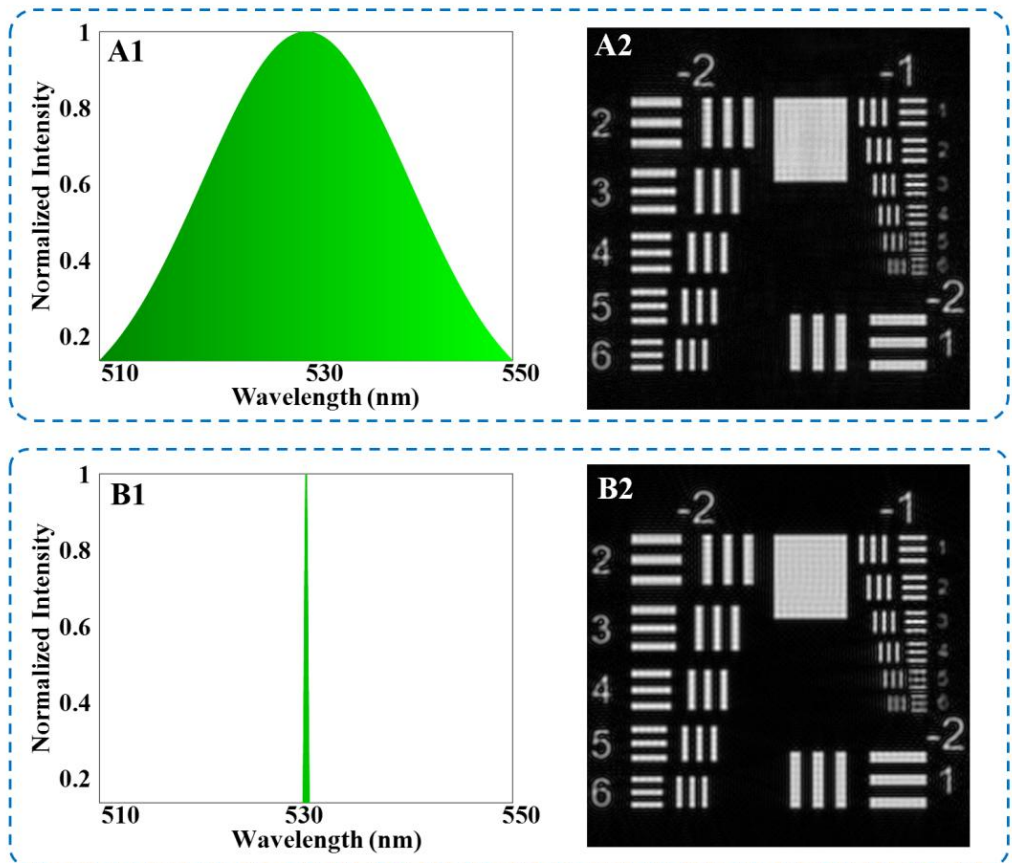

**Fig. S3. The effect of incoherent LEDs on reconstruction images.** (A1) The wide spectrum of LEDs; (A2) The reconstruction image illuminated by LEDs with wide spectrum. (B1) The narrow spectrum of LEDs; (B2) The reconstruction image illuminated by LEDs with narrow spectrum.

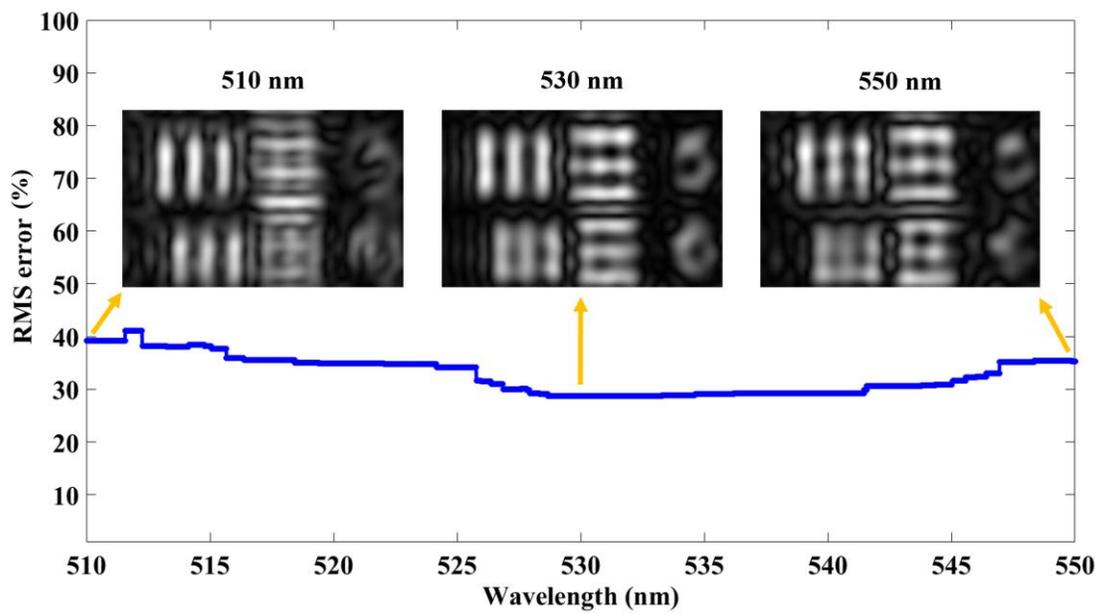

**Fig. S4. The RMS error of mini-FEI super-resolution imaging chip at different illumination wavelengths.**

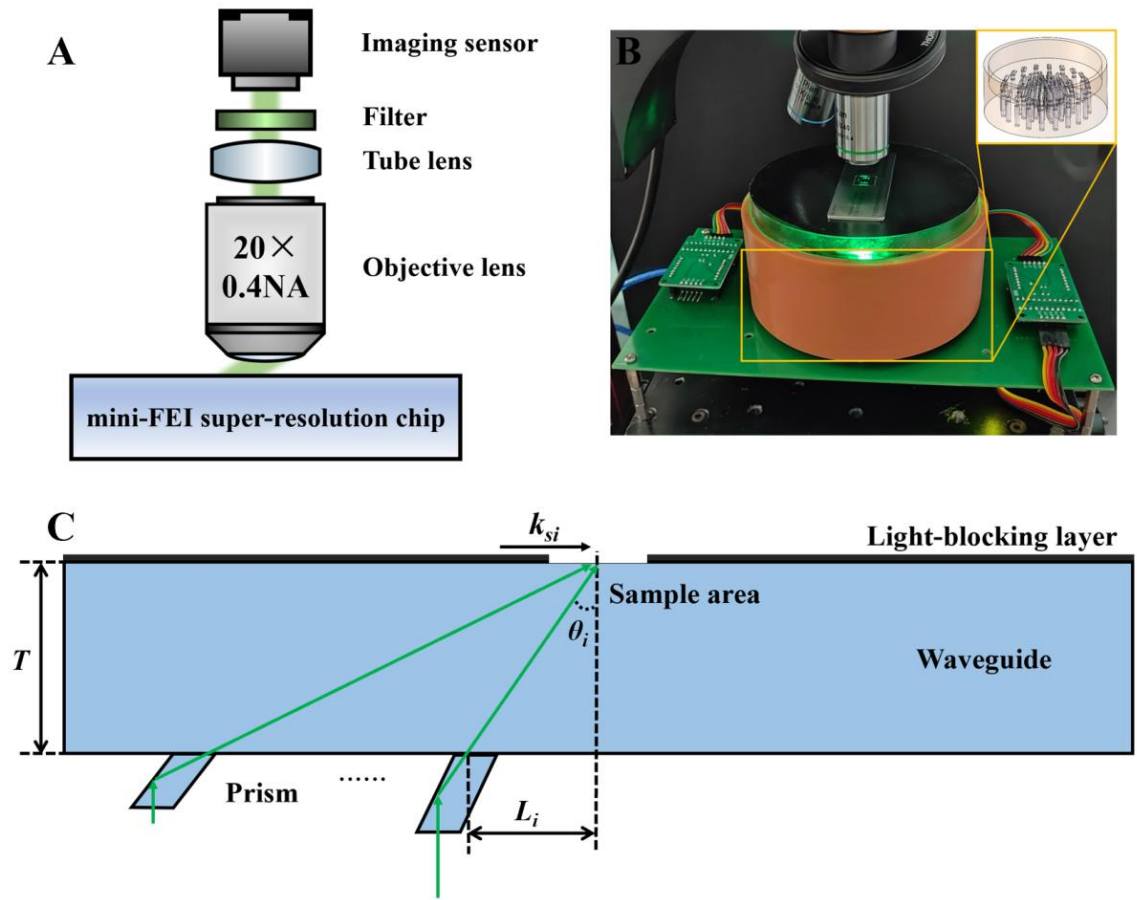

**Fig. S5. The mini-FEI super-resolution imaging system.** (A) The schematic of mini-FEI super-resolution microscopic system. (B) The physical diagram of mini-FEI super-resolution imaging chip. (C) The cross-section view of mini-FEI super-resolution imaging chip.

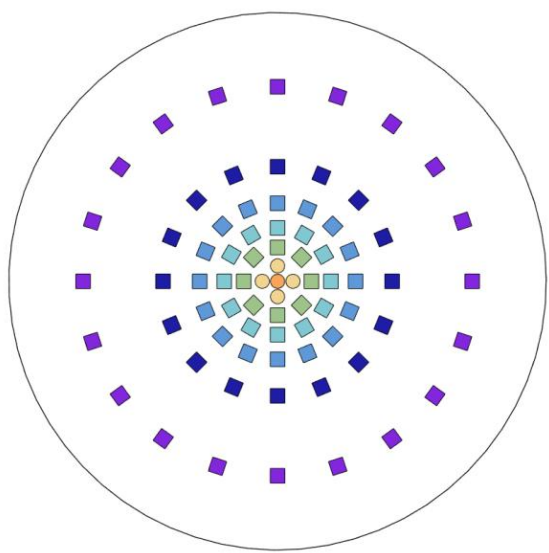

**Fig. S6. The schematic of prism distribution.**

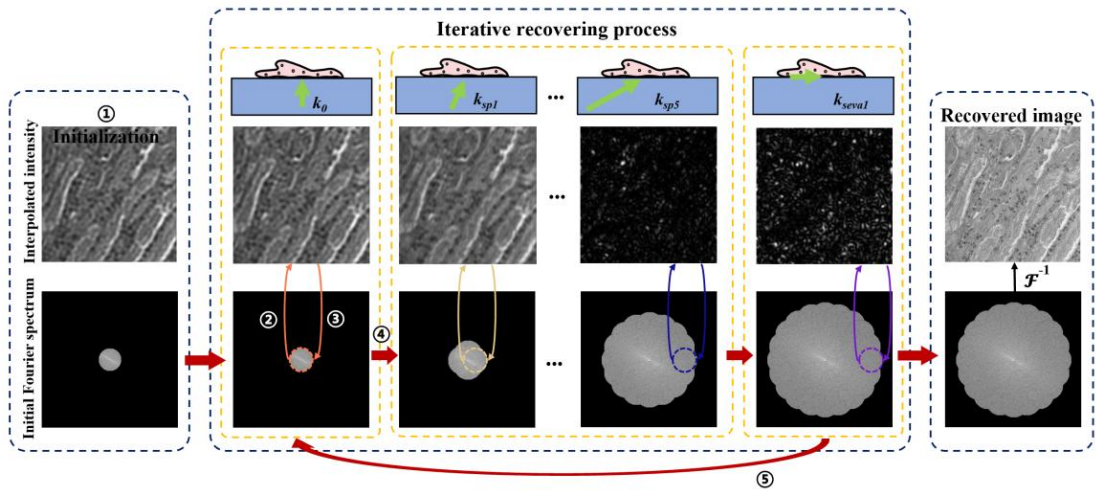

**Fig. S7.** The recovery algorithm flow chart of mini-FEI super-resolution imaging.

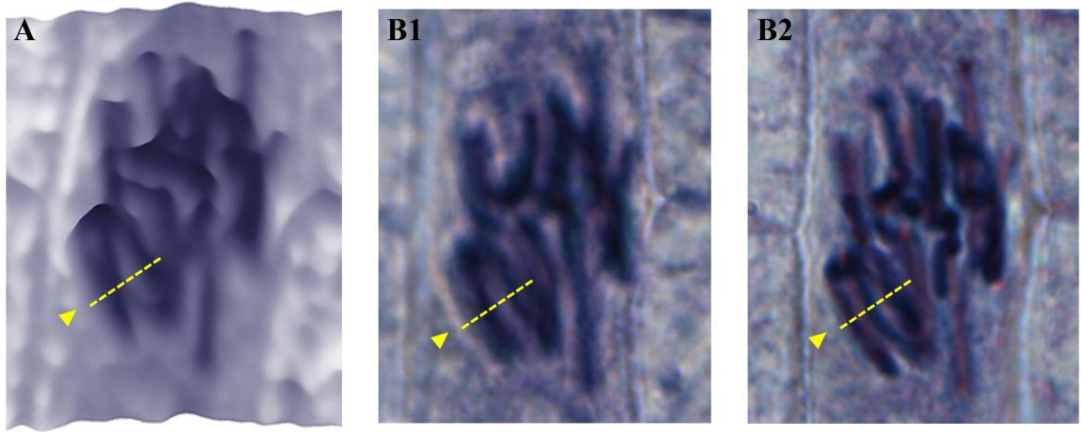

**Fig. S8. 3D reconstruction images containing information of different focus planes.** (A) The 3D reconstruction image of metaphase of mitosis stage in the root tip cells of onions. (B) The GT images of metaphase of mitosis stage at different focus planes.

**Supplementary Table 1** The parameters of mini-FEI super-resolution imaging chip

| Circles | Incident angle/° | $n_{eff}$ | Overlap rate | Numbers of the prism | $L$/mm |
|---|---|---|---|---|---|
| 1 | 6 | 0.16 | 75% | 4 | 3.15 |
| 2 | 13 | 0.34 | 72% | 8 | 6.93 |
| 3 | 20 | 0.52 | 72% | 12 | 10.92 |
| 4 | 28 | 0.72 | 69% | 16 | 15.95 |
| 5 | 38 | 0.94 | 65% | 16 | 23.44 |
| 6 | 53 | 1.22 | 56% | 20 | 39.81 |

**Supplementary Table 2** Comparison of super-resolution imaging technologies.

| Papers | Sample type | Space Bandwidth Product(SBP) | Fabrication processing | Principle based |
|---|---|---|---|---|
| Phys.Rev.Lett.(2017) | Label-free | $4\times10^5$ | Microfabrication | SFS |
| Adv. Funct. Mater.(2019) | Label-free | $3\times10^6$ | Direct laser writing | SFS |
| Nat.Photon.(2020) | Labelled | $3\times10^4$ | Lithography | SFS |
| Nat.Commun.(2021) | Labelled | $2\times10^4$ | Lithography | SFS |
| Adv. Sci.(2022) | Labelled/Label-free | $1\times10^5$ | Lithography/FIB | SFS |
| This work | Label-free | $3.4\times10^7$ | Without Microfabrication | SFS |